\documentclass[
    reprint,
    aps,
    pra,
    amsmath,
    amssymb
]{revtex4-2}

\usepackage[english]{babel}
\usepackage{csquotes}

\usepackage{mathtools}
\usepackage{siunitx}

\usepackage{graphicx}

\usepackage[
    colorlinks=true,
    linkcolor=blue,
    citecolor=blue,
    urlcolor=blue
]{hyperref}

\newcommand{\panelref}[2]{%
Fig.~\hyperref[#1]{\ref*{#1}(#2)}%
}
\newcommand{\panelpair}[3]{%
Figs.~%
\hyperref[#1]{\ref*{#1}(#2)}%
\ and\ %
\hyperref[#1]{\ref*{#1}(#3)}%
}
\newcommand{\panelrange}[3]{%
Fig.~\hyperref[#1]{\ref*{#1}(#2)--(#3)}%
}

\newcommand{\uu}[1]{\underline{\underline{#1}}}

\makeatletter
\renewcommand{\fnum@figure}{\textbf{Figure~\thefigure}}
\renewcommand{\fnum@table}{\textbf{Table~\thetable}}
\makeatother

\usepackage{bm}

\begin{document}

\title{Multistability and state-switching in series-coupled resonant tunneling diodes}

\author{Jannis Waldmann}
\email{jannis.waldmann@tu-ilmenau.de}

\author{Jonnel Jaurigue}
\author{Kathy Lüdge}

\affiliation{Institute of Physics, Technische Universität Ilmenau, 98693 Ilmenau, Germany}

\date{\today}

\begin{abstract}
    Resonant tunneling diodes (RTDs) embedded in an electrical circuit are known for their neuron-like response characteristics, which makes them promising candidates for neuromorphic applications. This paper investigates the dynamical response of series-coupled  RTDs and systematically analyzes the impact of coupling and inhomogeneities on the solution structure. We further propose a scheme for controlled switching between coexisting stable states which allows to realize tunable memory elements in these circuits.  The coupled RTD system exhibits a rich bifurcation structure, showing different degrees of multistability between symmetric and antisymmetric solutions. Limit-cycle branches and their dependence on the system parameters are analyzed using numerical continuation methods. A central focus is placed on the role of symmetry. For two identical RTDs, the system possesses a $\mathbb{Z}_2$ exchange symmetry, which governs the emergence of symmetry-breaking bifurcations and multistable states. The analysis is further generalized to $N$ coupled RTDs, revealing the underlying $S_N$ symmetry structure and its influence on the organization of equilibrium branches, paving the way for neuromorphic network operation.
\end{abstract}

\maketitle

\section{INTRODUCTION}\label{sec:introduction}
Dynamical mechanisms underlying information processing in neural systems are not yet fully understood. At the same time, the search for energy-efficient hardware architectures has given rise to the development of neuromorphic computing concepts that emulate neuronal dynamics using physical devices \cite{KUD25}. To increase our understanding of neural system dynamics and further neuromorphic hardware development our present study focuses on the dynamic analysis of a circuit-level model of coupled resonant tunneling diodes (RTDs) \cite{IRO23}. From this analysis of coupled RTDs we demonstrate new neuromorphic computer functionalities, particularly multi-state switching and operations as memory elements.

This work devoted attention to physical systems, specifically RTDs, that naturally exhibit neuron-like dynamics. This is because although mathematical descriptions of neuronal activity ranging from detailed biophysical models such as the Hodgkin-Huxley model \cite{HOD52,HOD52a} to reduced descriptions including the FitzHugh-Nagumo model \cite{FIT55,CEB24} and the Izhikevich model \cite{IZH03} provide important theoretical insight, excitable physical systems and networks of coupled oscillators continue to play a fundamental role for future physical computing hardware \cite{TOD24,ROM23,ROM17}. Initial descriptions of RTD-related tunneling in finite superlattices were provided by Tsu and Esaki in 1973 \cite{TSU73}, while the measurement of the resonant tunneling mechanism was first conducted in 1974 by Chang et al.~\cite{CHA74}. The initial RTD design was proposed in \cite{SOE89}, which denotes a layered structure comprising disparate semiconductors. From here, production processes and materials for RTDs continued development \cite{CIM22,SAM23}.

We analysed RTDs because they exhibit intricate dynamic characteristics, including hysteresis and bistability \cite{LIU88, TEI19} as well as Andronov-Hopf bifurcations and slow-fast dynamics \cite{WAL93, ORT21, ORT21a}. The corresponding limit cycle solutions are canard solutions \cite{WEC13,ORT22}. Responses to external input, such as pulses, were also investigated \cite{ROM17} in view of the fact that RTDs exhibit phenomena including excitable spiking \cite{ROM13a,JAC25,ZHA21}, bursting \cite{ORT21, ROM17} and integrate-and-fire mechanisms \cite{ROB24b}. Recent investigations have led to the observation that single RTDs also exhibit a spiking flip-flop memory \cite{DON24c}, meaning that the RTD can switch between a stable state and stable oscillating behavior, which can be controlled by induced external voltage pulses. These dynamic features of excitable RTDs have been intensively exploited in the context of spiking neural networks \cite{ROM23, OWE26, DON25b, HEJ24,HEJ22} and they are also promising for applications in the frameworks of spatially or time-multiplexed reservoir computing \cite{ABB25a,MUE24,DON25}. 

In addition to their intricate dynamics, RTDs are candidates for the fastest electronic oscillators with a reported operational frequency of $\qty{1.98}{\tera\hertz}$ \cite{IZU17}. This advancement makes them suitable for utilization in wireless communication systems and other applications where speed is important. Due to their small dimensions \cite{WAN15a,OSH16, OSH16a}, the components can be built into a programmable neuromorphic integrated chip as, e.g., done by TrueNorth \cite{MER14a}  or Intel Quark SE \cite{IQ16}. This integration results in mechanisms of information transmission similar to those in the human brain and nervous system.

With these intricate dynamics and promising hardware applications of RTDs in mind, we build on past RTD research with our systematic bifurcation analysis of coupled RTDs. In Sec.~\ref{sec:model} we will introduce RTDs themselves, as well as the circuit and the equations that model the dynamics of the series-coupled RTDs.

Sec.~\ref{sec:dynamics} provides an overview over the fixed-point and limit cycle branches of the series-coupled RTDs with respect to the state-switching property. This state-switching property follows on from our current understanding of networks of coupled dynamical systems. It is known that their rich dynamics are due to the interaction between the different nodes which increases the level of multistability \cite{ROE18,CAM01,SHI18} and can lead to effects like amplitude death \cite{ILL16}, transient chaos \cite{SAN21} or temporal localized states \cite{MAR24b}.  Additionally, as shown in \cite{HEI10} coupled nonlinear circuit elements with cubic nonlinearities lead to multistability with an underlying symmetry structure. We provide guidelines for suitable parameter regions generating this state-switching property. Occurring bifurcations are also identified and discussed in dependence of the system's parameters. In this context, tipping points may occur, marking qualitative transitions between distinct dynamical states \cite{SCH09,AMB21}.

We further investigate the state switching property of coupled RTDs in Sec.~\ref{sec:state_switching}. We demonstrate the controlled switching of stable states that can be achieved by applying an external current input to the circuit. We then describe operation guidelines for coupled RTDs that exploit this property for innovative neuromorphic applications, and discuss the limit of $N$ coupled RTD units.

\section{COUPLED RTD MODEL}\label{sec:model}
\begin{figure}[ht]
    \centering
    \includegraphics[width=\linewidth]{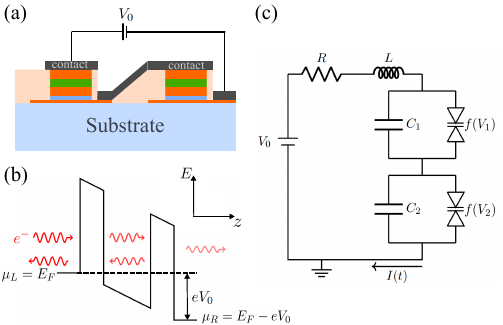}
    \caption{(a) Exemplary layer structure for the series-coupled RTD circuit with diagonal contact bar. Orange and green regions indicate differently doped semiconductor layers. (b) Potential landscape of a single RTD double-barrier quantum well with applied voltage $V_0$ and chemical potentials $\mu_L$ and $\mu_R$ on both sides. (c) Schematic equivalent circuit of two series-coupled RTD oscillators with nonlinearity $f(V_{1/2})$.}
    \label{fig:Panel1}
\end{figure}

\begin{figure}[ht]
    \centering
    \includegraphics[width=\linewidth]{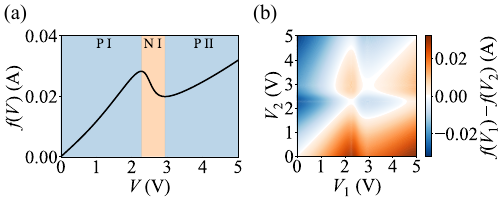}
    \caption{(a) Nonlinear current-voltage characteristics of a single RTD (black line), regions of positive/negative differential conductance (blue/orange shading) with label P/N. (b) Critical manifold $\mathcal{C}_0$ (Eq.\eqref{eq:C0}) for the series coupled model.}
    \label{fig:Panel2}
\end{figure}

The simplest configuration for two RTDS is a series-shunted model in which the two RTDs are coupled with each other and both are driven by a single voltage source as in \panelref{fig:Panel1}{c}. 
\noindent
An exemplary layer structure for this is given in \panelref{fig:Panel1}{a}, wherein both RTDs are represented as layered semiconductor heterostructures embedded in a substrate and contacted by conducting bars. Green bars in the RTD layers represent narrow-band-gap semiconductor regions sandwiched between layers of a wider-band-gap semiconductor (see \cite{IRO23} for more details on the RTD itself). As is evident from the configuration, the potential of the RTD forms a double-barrier quantum-well (DBQW) as shown in \panelref{fig:Panel1}{b}. Owing to the interference of incident and reflected electron waves, bound quantum states can be established within the well. This interference assists the tunneling of incident electrons, which can be observed in the nonlinear and non-monotonic current-voltage relation, given in \panelref{fig:Panel2}{a}. The applied bias voltage $V_0$ acts as a linear shift on the DBQW potential. The energy $eV_0$ results in the occurrence of two distinct chemical potentials on the left side, designated as $\mu_L$ in \panelref{fig:Panel1}{b}, and on the right side, denoted as $\mu_R$, of the DBQW, respectively. Following \cite{SCH96o}, the RTD's current-voltage equation can be derived by applying the Tsu-Esaki equation to the Lorentzian transmission coefficient of the DBQW, under the assumption of Fermi-Dirac statistics for the charge carriers. An additional diode-like exponential term ensures the increasing valley current for higher voltages.
\begin{flalign}\label{eq:RTD_curr_vol}
    f(V) = &a\ln\left\{\frac{1 + \exp[(e/k_B T)(b-c+n_1 V)]}{1 + \exp[(e/k_B T)(b-c-n_1V)]}\right\} && \notag \\
    &\times \left\{\frac{\pi}{2} + \arctan\left(\frac{c-n_1V}{d}\right)\right\}\\
    &+ h\left\{\exp[(e/k_B T)n_2 V] - 1\right\} && \notag
\end{flalign}
In Eq.~\eqref{eq:RTD_curr_vol}, $T$ is the temperature, $e$ is the elementary charge and $k_B$ the Boltzmann constant. The used parameters are $a=\qty{0.0039}{\ampere}$, $b=\qty{0.05}{\volt}$, $c=\qty{0.0874}{\volt}$, $d=\qty{0.0073}{\volt}$, $n_1=0.0352$, $n_2=0.0031$, $h=\qty{0.0367}{\ampere}$, $T=\qty{300}{\kelvin}$ according to \cite{SCH96o}. These parameters represent the structural properties of the RTD, including materials, dimensions and charge carrier density and yield a smooth nonlinear current voltage curve, depicted in \panelref{fig:Panel2}{a}. The given set of parameters corresponds to the typical N-shaped curve for III-V semiconductor RTDs, which shows a region of negative (N) differential conductance between two regions of positive (P) differential conductance.
\noindent
It is noted that other parameter sets, e.g.~those used in \cite{OWE26}, show more abrupt transitions between N and P regions. There are similarities to the idealized cubic nonlinearity of the FitzHugh–Nagumo system that are widely used to model the dynamic response of neurons, however the RTD nonlinearity is based on existing physical devices which leads to the more complex functional dependencies. The equivalent circuit of series-coupled RTDs is depicted in \panelref{fig:Panel1}{c}. It comprises two series-shunted RTDs arranged parallel with parasitic capacities $C_{1,2}$ and in series to a voltage source $V_0$, a resistor $R$ and an inductor $L$. The capacities are attributed to the charged layers that are formed on both sides of the DBQW as well as the resistance and inductance resulting from the connection to cables and contact bars within the layer scheme. The dynamics of the circuit are described by three coupled first-order differential equations, as was similarly proposed in the literature \cite{LIU88, WAL93}. These equations are based on the Kirchhoff laws for a single RTD powered by static bias voltage, but now extended for the equivalent circuit, given in \panelref{fig:Panel1}{c}. There are also large-signal models for RTDs, as e.g.~given in \cite{OUR23}, but due to their purely numerical description of the current-voltage characteristics they are not suitable for our dynamical investigations.
\begin{flalign}
    \dot{V}_1 &= \mu^{-1} (I - f_1(V_1)) \label{eq:fir_RTD}\\
    \dot{V}_2 &= (\mu \kappa)^{-1} (I - f_2(V_2)) \label{eq:sec_RTD} \\
    \dot{I} &= \mu (V_0 - V_1 - V_2 - RI) \label{eq:thi_RTD}
\end{flalign}
The additional Eq.~\eqref{eq:sec_RTD} is attributed to the second RTD, consequently enabling the interpretation of the voltages $V_{1,2}(t)$ as those that drop over the first and second RTD-capacity combination while the current $I(t)$ is the total current within the single loop. In this set of equations, the time is scaled with the dimensionless time $t=\omega_{10} \Tilde{t}$, according to \cite{ORT21}. Therein, $\omega_{10} = 1/\sqrt{LC_1}$ is the first RTD's natural frequency and $\mu=\sqrt{C_1/L}$ is the reduced parameter acting as the stiffness parameter of the system. Since the capacity of the second RTD, $C_2$, may differ with respect to $C_1$, another parameter is $\kappa=C_2/C_1$. 
In the context of identical RTDs, characterized by the condition that $f_1(V_1) = f_2(V_2)$ and $\kappa=1$, the system manifests a $\mathbb{Z}_2$ exchange symmetry between the voltages $\mathbb{Z}_2:(V_1,V_2)\to(V_2,V_1)$. \\
For $\mu \ll 1$, the system exhibits a slow-fast structure \cite{ORT21}, where the voltages $V_1$ and $V_2$ evolve on a fast timescale while the current $I$ changes slowly. Introducing the fast time scale $\tau = t/\mu$ and considering the singular limit $\mu\to 0$, the fast voltage variables rapidly relax to their quasi-equilibrium states for a frozen value of the slow variable $I$. This adiabatic elimination of the voltage degrees of freedom defines the following critical manifold $\mathcal{C}_0$,
\begin{align}
    \mathcal{C}_0 = \left\{(V_1, V_2, I): I = f(V_1) = f(V_2)\right\}.\label{eq:C0}
\end{align}
For $f_1 = f_2\equiv f$, the set can be visualized by the zeros of $F(V_1,V_2)=f(V_1) - f(V_2)$, i.e., the curve along which the solutions of Eqs.~\eqref{eq:fir_RTD}-\eqref{eq:thi_RTD} lie, which can be seen in \panelref{fig:Panel2}{b} as white region. The critical manifold reveals a symmetric center branch along the diagonal $V_1 = V_2$, as well as two antisymmetric branches (wing-like curves emerging from the diagonal) which can be transformed into one another via the $\mathbb{Z}_2$ symmetry. Along the antisymmetric branches, both RTDs operate at different voltages.
\section{DYNAMICAL ANALYSIS}\label{sec:dynamics}
The following analysis considers the specific case $f_1(V_1)=f_2(V_2)$, meaning that the two RTDs are described by the same current-voltage curve. Nevertheless, $\kappa$ remains as a free variable.

\begin{figure}[ht]
    \centering
    \includegraphics[width=\linewidth]{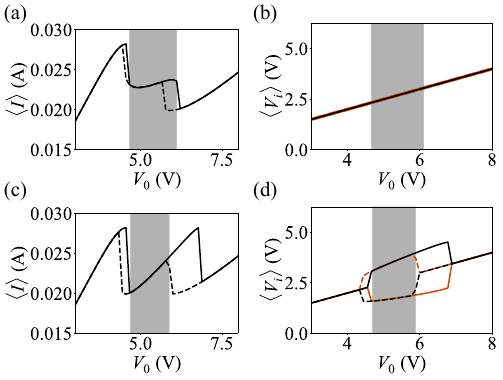}
    \caption{Hysteresis scans of the series-coupled RTD system for $\kappa=1$ (top row) and $\kappa=1.1$ (bottom row) showing the mean values $\langle V \rangle$ and $\langle I \rangle$. In (a) and (c), black solid/dashed curves denote forward/backward scans. Grey shadings indicate oscillating regions. In (b) and (d), black solid/dashed curves indicate the forward/backward scans of the first RTD, while orange solid/dashed curves indicate the corresponding scans of the second RTD.}
    \label{fig:Panel3.1}
\end{figure}

As previously proposed in \cite{WAL93,HEI10}, circuits containing nonlinear elements, in particular single-RTD circuits, exhibit hysteretic behavior. In such systems, upward and downward scans of the bias voltage $V_0$ lead to different dynamical responses. The corresponding scans,  where the mean voltages $\langle V_i \rangle$ and the mean current $\langle I \rangle$ are recorded  after an initial transient, are shown in Fig.~\ref{fig:Panel3.1} with identical RTDs ($\kappa = 1$) in panel (a,b)  and non-identical series-coupled RTDs ($\kappa = 1.1$) in panel (c,d). The regions where the voltages oscillate are indicated with grey shadings.
Black curves correspond to $\langle V_1 \rangle$ and orange curves to $\langle V_2 \rangle$. Furthermore, solid and dashed lines indicate upward and downward scans, respectively.

For $\kappa = 1$, the system exhibits the characteristic current hysteresis known from a single RTD. The voltages, however, evolve along the symmetric branch and do not show hysteresis; hysteretic behavior is observed only in the current. For $\kappa = 1.1$, the current characteristics change significantly, and hysteresis becomes apparent in both the current and the voltages (second row in Fig.~\ref{fig:Panel3.1}).
\noindent
To better explain the observed hysteresis behaviour, the following section will now discuss the solution structure of the coupled RTD system in more detail.
Equilibrium branches and continued bifurcations were obtained with BifurcationKit \cite{VEL20} in Julia while time series and phase-space trajectories were obtained by Python simulations.

\subsection{Fixed points}
Fixed points can be found as equilibria of the dynamical system where the nullclines of the equations intersect, resulting in the following two equations.
\begin{flalign}
    I &= f(V_1) = f(V_2) \label{eq:fixed_points1}\\
    V_0 &= V_1 + V_2 + RI \label{eq:fixed_points2}
\end{flalign}
This set of equations describes in fact the earlier defined critical manifold $\mathcal{C}_0$ with the additional condition $V_0 = V_1 + V_2 + RI$. The stability of the fixed points is determined by the real part of the eigenvalues of the Jacobian of Eqs.~\eqref{eq:fir_RTD}-\eqref{eq:thi_RTD}.
\begin{gather}
    \uu{\bm{J}}  = \left(\begin{array}{ccc}
            -\frac{f_1^{(1)}(V_1)}{\mu} & 0 & \mu^{-1}\\
            0 & -\frac{f_2^{(1)}(V_2)}{\mu \kappa} & (\kappa\mu)^{-1} \\
            -\mu & -\mu & -\mu R
        \end{array}\right)
\end{gather}
Here, $f^{(n)}_i(X)|_{X=X_P}$ denotes $\frac{d^n f_i(X)}{dX^n}\bigr|_{X=X_P}$ with a variable $X\in\mathbb{R}$ of the $i$-th RTD and $n\!\in\!\mathbb{N}_0$. For $\kappa=1$ and $f_1(V_1)=f_2(V_2)$, the exchange symmetry $\mathbb{Z}_2$ implies that the Jacobian can be decomposed into symmetric and antisymmetric eigenmodes, leading to a simplified eigenvalue structure. A more thorough discussion of the eigenvalues and their dependence on $V_0$ can be found in Sec.~\ref{sec:bifurcations} and in the Appendix, see Sec.~\ref{sec:evalues}.

\subsection{Equilibrium branches, bifurcations and oscillations}\label{sec:bifurcations}

\begin{figure*}[t]
    \centering
    \includegraphics[width=\textwidth]{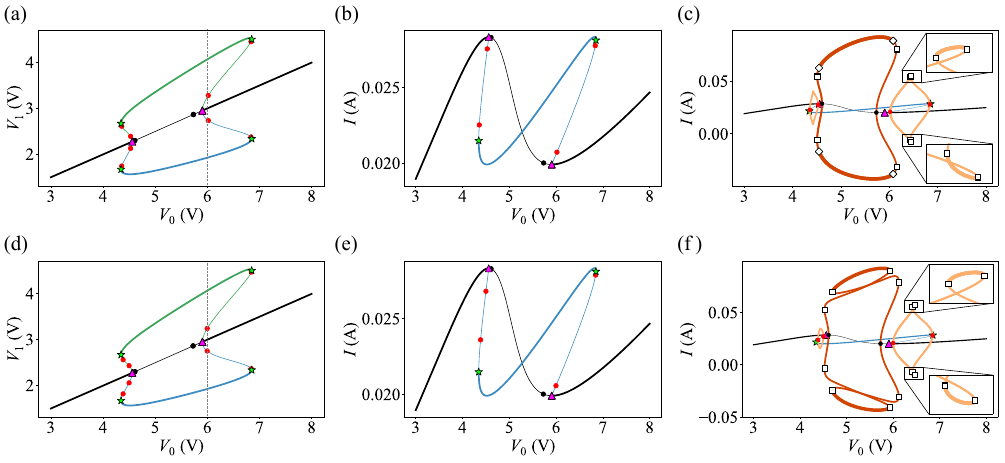}
    \caption{Bifurcation diagrams for two series coupled RTDs. Thick/thin lines correspond to stable/unstable  solutions. Symbols mark bifurcation points. Red and black dots indicate Andronov-Hopf bifurcations, magenta triangles subcritical Pitchfork bifurcations and green stars saddle-node bifurcations. White squares indicate saddle-node bifurcations of limit cycles and white diamonds branch points of limit cycles. (a) Voltage fixed-point branches. Black lines indicate symmetric and blue and green lines antisymmetric fixed point branches. (b) Current fixed-point branches. In this projection, the antisymmetric blue branch overlaps the green branch. (c) Current fixed-point branches in blue and black together with continued limit cycles oscillation branches. Therein red limit cycles correspond to symmetric oscillations and orange limit cycles to antisymmetric oscillations. (d)-(f) show the same bifurcation diagrams as (a)-(c) but for $\kappa = 1.1$. Dashed vertical grey lines in (a) and (d) indicate the bias voltage for the state switching shown in Sec.~\ref{sec:state_switching}.}
    \label{fig:Panel3}
\end{figure*}

Figure \ref{fig:Panel3} summarizes the bifurcation diagrams for the series-coupled RTDs. As can be seen in \panelref{fig:Panel3}{a}, the  bifurcation diagram in dependence of $V_0$ exhibits symmetrical (black) branches for $V_1 = V_2$ and antisymmetric fixed-point branches (green, blue) which are described by Eqs.~\eqref{eq:fixed_points1} and \eqref{eq:fixed_points2}. The appearance can be attributed to the fact that the set of fixed points is a projection of the critical manifold $\mathcal{C}_0$. In a similar manner, the two antisymmetric branches can be transformed into one another by applying the $\mathbb{Z}_2$ symmetry. The two antisymmetric branches separate from the symmetrical branch in subcritical pitchfork bifurcations (magenta triangles in \panelref{fig:Panel3}{a}).
\noindent
This is also confirmed by the eigenvalues, as the subcritical pitchfork corresponds to a single zero-crossing eigenvalue. To describe the splitting in the $V_0$ bifurcation diagram, the coordinates can be split into symmetric $s$ and antisymmetric $a$ coordinates.
\begin{gather}
    s = \frac{V_1 + V_2}{2},\ a = \frac{V_1 - V_2}{2}
\end{gather}
The application of the $\mathbb{Z}_2$ symmetry to the new coordinates yields the following transformations: $\mathbb{Z}_2:s\to s$ and $\mathbb{Z}_2:a\to -a$. It can be observed that symmetric solutions correspond to $a=0$, while antisymmetric solutions correspond to $a \gtrless 0$. Eqs.~\eqref{eq:fir_RTD}-\eqref{eq:thi_RTD} can now be transformed into the new coordinates $(s, a, I)$, yielding the following equations.
\begin{flalign}
    \dot{s} &= \frac{1}{2\mu}[2I - f(s+a) - f(s-a)] \label{eq:newcoord1}\\
    \dot{a} &= -\frac{1}{2\mu}[f(s+a)-f(s-a)] \label{eq:newcoord2}\\
    \dot{I} &= \mu (V_0 - 2s - RI) \label{eq:newcoord3}
\end{flalign}

In order to derive the normal form of the system in the proximity of the pitchfork bifurcation at $V_0 = V_{0,p}$, it is necessary to expand the right side of Eq.~\eqref{eq:newcoord2} for small perturbations. This is due to the fact that the pitchfork is caused by an instability in the antisymmetric direction. The resulting Taylor series simplifies because even exponentials cancel out, which results in the normal form of a pitchfork bifurcation \cite{STR15,KUZ23}.
\begin{flalign}
    \dot{a} \approx ra + \beta a^3 + \gamma a^5\label{eq:pitchfork}
\end{flalign}
The coefficients are $r=-\mu^{-1}\frac{df}{ds}\bigr|_{s=s_p}$, $\beta = - (6\mu)^{-1}\frac{d^3f}{ds^3}\bigr|_{s=s_p}$ and $\gamma = -(120\mu)^{-1}\frac{d^5f}{ds^5}\bigr|_{s=s_p}$ with $s_p = s(V_{0,p})$. The condition for the occurrence of the pitchfork is that the coefficient of the linear term $r$ undergoes a change in sign and therefore the condition follows to $f^{(1)}(s)|_{s=s_p} = 0$. Consequently, the first pitchfork bifurcation occurs at the local maximum of the RTD current–voltage function (see \panelref{fig:Panel3}{b}) which is consistent with our numerical path continuation results. Because the stiffness parameter $\mu > 0$, the coefficient of the cubic term determines whether the pitchfork is supercritical $f^{(3)}(s)|_ {s=s_p}>0$ or subcritical $f^{(3)}(s)|_{s=s_p}<0$ and requires $f^{(3)}(s)|_{s=s_p} \neq 0$. $\gamma < 0$, meaning $f^{(5)}(s)|_{s=s_p}>0$ provides the stabilizing quintic term. Consequently the quintic term opposes the growth of $|a|$ and bounds the dynamics of the symmetry-broken solutions.

Saddle-node bifurcations appear on the antisymmetric branches on the outer left and right side (see green stars on the green and blue lines in \panelref{fig:Panel3}{a}). These bifurcations can be described in two ways: (1) within the framework of the series expansion for the pitchfork since the saddle-node bifurcation emerges from the pitchfork or (2) in a more global understanding. In the first approach, the series expansion of the pitchfork is employed and the equilibria of the reduced dynamics are considered, where non-trivial ones satisfy the following expression.
\begin{gather}
    0 = r + \beta a^2 + \gamma a^4
\end{gather}
The saddle-node bifurcation occurs when two equilibria coalesce, which is characterized by a zero of the first derivative of Eq.~\eqref{eq:pitchfork}. This corresponds to the points at which an eigenvalue of the Jacobian of the normal form crosses zero.
\begin{gather}
    0 = r + 3\beta a^2 + 5\gamma a^4 
\end{gather}
The subtraction of both conditions yields
$0 = 2\beta a^2 + 4\gamma a^4$ which admits the non-trivial solution
\begin{gather}
    a^2 = -\frac{\beta}{2\gamma} = -\frac{10 f^{(3)}(s)|_{s=s_p}}{f^{(5)}(s)|_{s=s_p}}.
\end{gather}
Here, $a$ denotes the amplitude at the saddle-node bifurcation of the reduced normal form. The coefficients $\beta$ and $\gamma$ remain evaluated at $s_p=s(V_{0,p})$, since the reduced dynamics are obtained from a Taylor expansion around the pitchfork bifurcation. 

For the second approach to find saddle-node bifurcations, we consider the second fixed-point equation (Eq.~\eqref{eq:fixed_points2}) and parametrize $V_2$ as function of $V_1$, i.e., $V_2=V_2(V_1)$.
\begin{gather}
    V_0 = V_1 + V_2(V_1) + Rf(V_1)
\end{gather}
The saddle-node bifurcation occurs when $dV_0/dV_1 = 0$. Using $\frac{dV_2}{dV_1} = \frac{f^{(1)}(V_1)}{f^{(1)}(V_2)}$, the derivative of $V_0$ with respect to $V_1$ becomes $dV_0/dV_1 = 1 + f^{(1)}(V_1)/f^{(1)}(V_2) + Rf^{(1)}(V_1)$. Consequently, the saddle-node bifurcation is characterized by the two following conditions.
\begin{flalign}
    0 &= 1 + \frac{f^{(1)}(V_1)}{f^{(1)}(V_2)} + Rf^{(1)}(V_1) \label{eq:fold1}\\
    f(V_1) &= f(V_2)\label{eq:fold2}
\end{flalign}
Eqs.~\eqref{eq:fold1} and \eqref{eq:fold2} describe the saddle-node bifurcation on the full equilibrium manifold, providing an exact characterization. In contrast, the first derivation is based on a local approximation of the antisymmetric dynamics and shows how the saddle-node emerges from the pitchfork bifurcation through higher-order nonlinearities. \\
The antisymmetric branches restore later again within a pitchfork bifurcation on the main branch because the RTD current-voltage curve has a second $f^{(1)}(s)=0$ point at the local minimum. Another projection of the bifurcation diagram, shown in \panelref{fig:Panel3}{b}, reveals that the antisymmetric branches exhibit the same current (only one blue branch is seen which lies on top of the green branch), which can again be explained by the $\mathbb{Z}_2$ exchange symmetry of the voltages resulting from the coupling in series.

Emergent Andronov-Hopf bifurcations in Fig.~\ref{fig:Panel3} are marked as black dots on the symmetric branch and red dots on the antisymmetric branches. Notably, for identical RTDs ($f_1(V_1) = f_2(V_2)$ and $\kappa=1$ discussed in the first row of Fig.~\ref{fig:Panel3}), the Andronov-Hopf bifurcations have similar coordinates on the bifurcation diagram as for the single RTD, but at twice the bias voltage. However, in the series-coupled system, the branch is already a saddle fixed point at the Andronov-Hopf bifurcation due to the subcritical pitchfork bifurcation in front of it. Andronov-Hopf bifurcations on the symmetric branch can be described in a symmetric subsystem $(s, I)$.
\begin{flalign}
    \dot{s} &= \frac{1}{\mu}[I-f(s)] \\
    \dot{I} &= \mu(V_0 - 2s - RI)
\end{flalign}
The characteristic polynomial of the Jacobian of the symmetric subsystem $\uu{\bm{J}}_{\text{sym}}$ is given by the following expression.
\begin{gather}
    \lambda^2 + \lambda \left(\frac{f^{(1)}(s)}{\mu} + \mu R\right) + Rf^{(1)}(s) + 2 = 0
\end{gather}
At the Andronov-Hopf bifurcation $V_0 = V_{0,\text{AH}}$, it holds $\text{tr}(\uu{\bm{J}}_{\text{sym}}|_{s=s_{\text{AH}}}) = 0$, $\text{det}(\uu{\bm{J}}_{\text{sym}}|_{s=s_{\text{AH}}})>0$ and $d(\text{tr}(\uu{\bm{J}}_{sym}|_{V_0 = V_{0,\text{AH}}}))/dV_0 \neq 0$ with $s_{\text{AH}}=s(V_{0,\text{AH}})$ as transversality condition \cite{STR15,KUZ23} leading to the following conditions.
\begin{gather}
    f^{(1)}(s)|_{s=s_{\text{AH}}} = -\mu^2 R\\
    2 - \mu^2 R^2 > 0
\end{gather}
Along the symmetric equilibrium branch $V_0 = 2s + Rf(s)$, the transversality condition reduces to $f^{(2)}(s)|_{s=s_{\text{AH}}} \neq 0$, which is satisfied. For a detailed derivation, see Sec.~\ref{sec:transversality_condition} in the Appendix. The antisymmetric branches cannot be described by a similar antisymmetric subsystem, therefore Andronov-Hopf bifurcations on these branches have to be analyzed using the full system.

Limit cycle branches can be continued from the Andronov-Hopf bifurcations, as seen in \panelref{fig:Panel3}{c} by the red and orange lines. Unlike the limit-cycle branches on the antisymmetric equilibrium branches, the limit-cycle branch on the symmetrical branch (red lines) shows the same characteristics as for the single RTD \cite{ORT21}. The symmetric limit-cycle branch becomes stable in a branch point of periodic orbits, indicated by white diamonds. The antisymmetric limit-cycle branch (orange lines) exhibits a loop and undergoes a saddle-node bifurcation.

The bifurcation diagrams in \panelrange{fig:Panel3}{d}{f} show the same scans as in \panelrange{fig:Panel3}{a}{c}, but for $\kappa=1.1$, which represents non-identical RTDs. Since the fixed points given by Eqs.~\eqref{eq:fixed_points1} and \eqref{eq:fixed_points2} are independent of $\kappa$, the equilibrium branches (black, blue and green lines in Fig.~\ref{fig:Panel3}) do not change; however the positions of the Andronov-Hopf bifurcations (filled circles) do change. It can also be noted that the limit-cycle branches (orange and red lines in \panelref{fig:Panel3}{f}) change drastically compared to \panelref{fig:Panel3}{c}, meaning that the stable range of the symmetric limit-cycle branch (red) decreases and the stable range of the antisymmetric limit-cycle branch (orange) increases when the RTDs become non-identical.

\subsection{Path continuation in 2D parameter space}
\begin{figure}[t]
    \centering
    \includegraphics[scale=1]{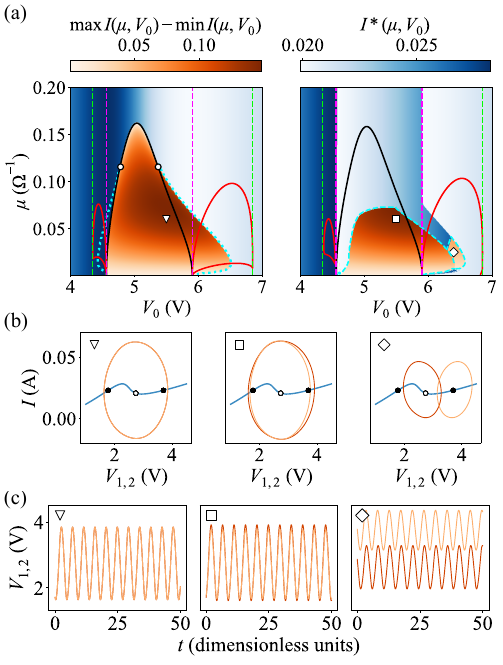}
    \caption{(a) 2D bifurcation scan for 2 coupled RTDs in a symmetric $\kappa=1$ (left) and antisymmetric $\kappa=1.1$ (right) arrangement obtained via upsweeping the voltage. Blue shadings indicate the current value at the stable fixed point and oscillating solutions are encoded via their oscillation amplitude (orange shadings). Black (red) lines indicate continued Andronov-Hopf bifurcations occurring on the symmetric (antisymmetric) branches, magenta curves pitchforks and green curves saddle-node bifurcations. White circles indicate generalized Andronov-Hopf points. Turquoise dashed lines indicate saddle-node bifurcations of limit cycles and turquoise dotted lines the stability change of the limit cycle branch for identical RTDs ($\kappa=1$, left). (b) Phase space representation of the limit cycles simulated at the respective signs in the 2D parameter space from (a). Red (orange) curves represent first (second) RTD, blue line shows $f(V)$ and black (white) circles indicate stable (unstable) fixed points. (c) Time series of the dynamics plotted in (b).}
    \label{fig:Panel4}
\end{figure}

The hysteretic behavior of the RTD system can be investigated further by performing two-dimensional parameter scans of its dynamical response. To this end, the bifurcations which were found in Fig.~\ref{fig:Panel3} can be continued in two-dimensional parameter space to gain a better understanding of the series-coupled RTD dynamics for different sets of parameters. The resulting bifurcation lines are shown in \panelref{fig:Panel4}{a} for identical ($\kappa=1$, left) and non-identical ($\kappa=1.1$, right) RTDs together with the results of direct numerical integration (orange and blue color code). 
For the latter, for each point on a $(i,j)\in(400\times 400)$ parameter grid, the system was integrated until the transient had decayed and the total current was evaluated. The scans were performed in direction of increasing bias voltage, such that the final state of the time series obtained for $V_{0, i-1}$ served as the initial condition for $V_{0,i}$. After completing one row in $V_0$, the parameter $\mu$ was incremented from $\mu_j$ to $\mu_{j+1}$ and the initial condition was set to $(V_1,V_2,I)=(0,0,0)$.
\noindent
Stable fixed-point solutions in \panelref{fig:Panel4}{a} are shown with blue shading, where the color indicates the fixed-point current. Oscillatory solutions are shown in orange, with the color representing the peak-to-peak current amplitude (time series of the symmetric oscillations can be seen in \panelpair{fig:Panel4}{b}{c} indicated with the triangle). Along the black(red) Andronov-Hopf bifurcation curves that occur on the symmetric(antisymmetric) branch, symmetric(antisymmetric) oscillations are born. Pitchfork bifurcations are indicated by vertical magenta dashed lines as given by Eq.~\eqref{eq:pitchfork}. Saddle-node bifurcations of fixed points are shown by vertical green dashed lines (they are given by Eqs.~\eqref{eq:fold1} and \eqref{eq:fold2}). The saddle-node bifurcations of limit cycles are obtained numerically and are indicated by turquoise dashed lines. 

In the case of $\kappa=1$ (left) in \panelref{fig:Panel4}{a}, the oscillating region forms a simply connected region in the $\mu$-$V_0$ parameter space. This oscillating region is partially bordered by the continued path of the Andronov-Hopf bifurcation  on the symmetric branch (solid black line in \panelref{fig:Panel4}{a}) and partially by the path of saddle-node bifurcations of limit cycles (turquoise dashed line in \panelref{fig:Panel4}{a}). The continued Andronov--Hopf bifurcation curve does not necessarily coincide with the boundary of the oscillatory region, as the stability of the limit cycles is determined by the saddle-node bifurcation curves of the limit cycles. Furthermore, an examination of the amplitude along the Andronov-Hopf curve reveals a transition from a supercritical to a subcritical Andronov-Hopf bifurcation, and subsequently back to a supercritical Andronov-Hopf bifurcation. The multistability between oscillating solutions and fixed points that we have discussed in the bifurcation scans in Fig.~\ref{fig:Panel3} gives rise to the hysteresis mentioned in Fig.~\ref{fig:Panel3.1}. For the 2D scans, this results in different dynamics observed numerically for different scan directions. The  turquoise dashed line seen to the left of the oscillating region is obtained during down-sweeping $V_0$ and marks the disappearance of oscillations via the saddle-node bifurcation of limit cycles (downsweep is not shown in \panelref{fig:Panel4}{a}). The turquoise dashed line on the right of the oscillating region is obtained via an up-sweep and directly borders the orange oscillating region. 

For $\kappa=1.1$, jumps in the fixed-point current are visible in the blue color scale. The oscillatory region is no longer a single connected region, but instead consists of two regions that share a common boundary. We note that the turquoise dashed lines indicate the saddle-node bifurcations of limit cycles originating from the symmetric and antisymmetric branches. In the positive $V_0$ direction, the system is on the symmetric fixed-point branch (black curve in \panelref{fig:Panel3}{f}). After switching from a stable node to saddle in the pitchfork bifurcation, the system jumps on the antisymmetric branch (blue curve in \panelref{fig:Panel3}{f}), due to the fact that the limit-cycle branch is not yet stable. This can be seen as the first jump in the fixed-point current. Once the limit-cycle branch (red curve in \panelref{fig:Panel3}{f}) becomes stable, the system dynamics changes to the symmetric limit-cycle branch, persisting until it becomes unstable in a saddle-node bifurcation of limit cycles. Due to the symmetric fixed-point branch's lack of stability (black curve in \panelref{fig:Panel3}{f}), the system transitions to the antisymmetric fixed-point branch (blue curve in \panelref{fig:Panel3}{f}) for high $\mu$ and for small $\mu$ on the stable limit-cycle branch on the antisymmetric branch (orange curve in the right part of \panelref{fig:Panel3}{f}). For high $\mu$, the system continues the antisymmetric fixed-point branch until the symmetric fixed-point branch stabilizes again in a pitchfork bifurcation. Thereafter, the system reverts to the symmetric branch. For smaller values of $\mu$, the system continues on the limit-cycle branch until it becomes unstable and switches finally back to the symmetric fixed-point branch.

Phase-space trajectories in \panelref{fig:Panel4}{b} and time series in \panelref{fig:Panel4}{c} show the dynamics on limit cycles corresponding to the selected points in parameter space in \panelref{fig:Panel4}{a} (first RTD plotted in red, second RTD in orange). It can be seen that, for identical RTDs ($\kappa=1$ labeled by the open triangle), both RTDs evolve along the same limit cycle. For non-identical RTDs ($\kappa=1.1$ labeled with the square), the symmetric limit cycle already results in slightly different dynamics of the two RTDs. On the asymmetric branch labeled by the open diamond, however, a qualitatively different behavior of the two RTDs (red and orange lines) is seen, supporting the interpretation of these branches as symmetric and antisymmetric.

\subsection{Symmetry and N-RTD generalization}

The underlying model can be readily extended to a system of $N$ series-coupled RTDs, each shunted by a parallel capacitance. The system is described by the state vector $\mathbf{S} = (V_1,\dots,V_N,I)\in\mathbb{R}^{N+1}$ and the capacitance ratios $\kappa_i = C_i/C_1$ and $\kappa_1 = 1$, where time is scaled with respect to the first RTD. For each additional RTD, a corresponding voltage equation is introduced, together with an associated term in the differential equation governing the current.
\begin{flalign}
    \frac{dV_i}{dt} &= \frac{1}{\mu \kappa_i}\Bigl(I - f_i(V_i)\Bigr), \quad i=1,\dots,N\\
    \frac{dI}{dt} &= \mu \left(V_0 - RI - \sum_{i=1}^N V_i\right)
\end{flalign}
For identical RTDs, i.e., $f_i(V_i) = f(V_i)$ and $\kappa_i = 1$, the system is equivariant under permutations of the RTD indices and therefore possesses $S_N$ symmetry.
\begin{figure}[ht]
    \centering
    \includegraphics[width=\linewidth]{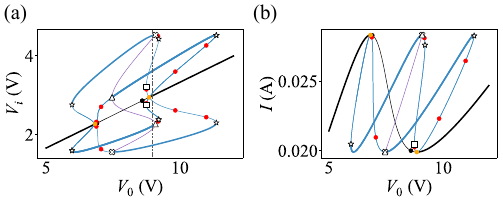}
    \caption{Bifurcation diagrams of fixed point solutions of 3 series-coupled RTDs as a function of $V_0$ showing solutions of the voltages $V_i$ (a) and the current I (b). Black/blue lines show symmetric/antisymmetric solutions, thick/thin lines encode stable/unstable solutions.  Symmetry-breaking bifurcation points of higher order are shown as orange dots, Andronov-Hopf bifurcations as red (antisymmetric) and black (symmetric) dots, and saddle-node/pitchfork bifurcations as white stars/white triangles. White crosses indicate other bifurcations. Vertical gray dashed line indicates a multistable state with 5 stable fixed points.}
    \label{fig:Panel4.1}
\end{figure}
\noindent
For $N=2$, this reduces to $S_2\cong\mathbb{Z}_2$, corresponding to the exchange symmetry of two devices. As already observed for $N=2$, the symmetry of the system determines the structure and multiplicity of equilibrium branches in the bifurcation diagram. Although the order of the symmetry group is $|S_N| = N!$, the relevant quantity is not the size of the group itself, but its action on the state space. In Fig.~\ref{fig:Panel4.1} bifurcation diagrams for the voltages $V_i$ and currents $I$ of $N=3$ RTDs are shown. The third RTD, compared to two series-coupled RTDs, leads to another antisymmetric branch as well as an unstable branch that connects antisymmetric branches with each other.

In particular, for $N$ RTDs, the state space decomposes into a one-dimensional symmetric subspace and an $(N-1)$-dimensional symmetry-breaking subspace. As $N$ increases, the number of possible symmetry-breaking configurations grows, leading to an increasing number of equilibrium branches which might be degenerate. This was also shown for different nonlinearities in \cite{HEI10}. Moreover, additional branches may emerge that connect equilibrium branches which are distinct for smaller values of $N$. Considering $N$ RTDs, the vector space of voltages $\mathbb{R}^N$ can be decomposed into a symmetric and a symmetry-breaking subspace, by defining a mean voltage $\bar{V} = \frac{1}{N}\sum_{i=1}^N V_i$ and the deviations from the mean $d_i = V_i - \bar{V}$. A basic property of the $d_i$ is
\begin{gather}
    \sum_{i=1}^N d_i = \sum_{i=1}^N (V_i - \bar{V}) = \sum_{i=1}^N V_i - N\bar{V} = 0.
\end{gather}
Consequently, the voltage space admits the decomposition
\begin{gather}
    \mathbb{R}^N = \text{span}(1,...,1) \oplus \left\{d\in\mathbb{R}^N: \sum_{i=1}^N d_i = 0\right\}.
\end{gather}
The first subspace corresponds to symmetric voltage configurations, whereas the second contains all symmetry-breaking perturbations. As shown previously for two RTDs, the symmetric subspace remains invariant under the action of $S_N$, while the symmetry-breaking subspace still carries the $N-1$ dimensional representation of $S_N$. From an experimental point of view, this means that adding RTDs leads to the emergence of stable antisymmetric states in which the system can be operated, while the symmetric state remains unchanged.

\section{STATE SWITCHING}\label{sec:state_switching}
\begin{figure}[t]
    \centering
    \includegraphics[width=\linewidth]{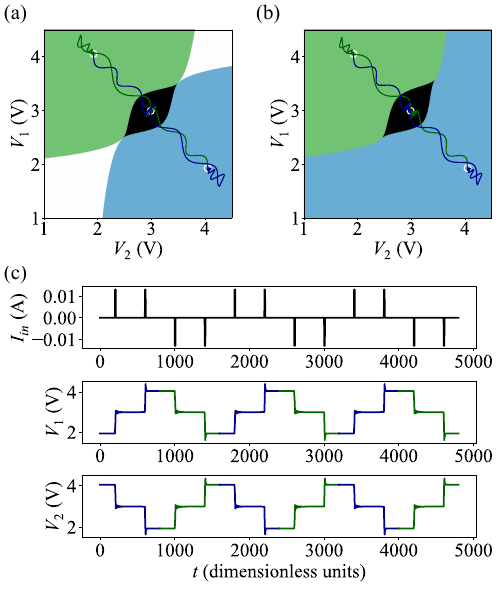}
    \caption{Attractor basins for stable fixed points for different $V_1$ and $V_2$ initial conditions for $\kappa=1$ in (a) and $\kappa=1.1$ in (b). The white region indicates oscillatory or divergent behavior. Fixed points are marked by dots with white outlines. Green, black, and blue regions correspond to initial conditions that converge to the fixed points of the respective color. Lines correspond to upward (cyan) and downward switching (lime) between fixed points. (c) Input current used to induce state switching (top panel) and corresponding voltage time series $V_{1,2}$ illustrating transitions between stable fixed points in the tristable regime (middle and bottom panels). Colors correspond to those used in panel (a) and (b).}
    \label{fig:Panel5}
\end{figure}

The bifurcation diagrams shown in \panelpair{fig:Panel3}{a}{b} reveal distinct regions of multistability. Of special interest are the tristable regimes located on the one hand for low voltages between the first saddle-node bifurcation on the antisymmetric branches (green stars) and the first pitchfork bifurcation on the symmetric branch (magenta triangle), and on the other hand for higher voltages between the second pitchfork bifurcation on the symmetric branch and the second saddle-node bifurcation on the antisymmetric branches. These regions suggest the possibility of controlled switching between coexisting states via external excitation. To this end, the series-coupled RTDs are biased with a voltage of $V_0 = \qty{6}{\volt}$, which lies within the second tristable regime. This is indicated by the vertical dashed grey line in \panelpair{fig:Panel3}{a}{d}. In \panelpair{fig:Panel5}{a}{b}, the basins of attraction for the different stable fixed points  are shown for parameter values of $R=\qty{1}{\ohm}$, $\mu = \qty{0.05}{\ohm}^{-1}$ and $V_0 = \qty{6}{\volt}$. The axes in \panelref{fig:Panel5}{a} correspond to initial conditions of the voltages, while the initial current is fixed at $I_0=\qty{0.028}{\ampere}$. The three stable fixed points that exist for the parameter choice are indicated by circles with white edges and distinct face colors. Consistent with the color of equilibrium branches in \panelref{fig:Panel3}{a}, blue and green correspond to antisymmetric and black to symmetric fixed points. 

\noindent

White regions in the case $\kappa=1$ indicate initial conditions for which trajectories either diverge or converge to a limit cycle; these regions disappear for $\kappa=1.1$ as there the parameter regions where stable oscillations are found shrink (compare bifurcation diagram in \panelpair{fig:Panel3}{c}{f}). Since a change in the bias voltage affects both voltages in a similar manner, it can already be seen that such control pulses would drive the system along the main diagonal in \panelpair{fig:Panel5}{a}{b}, which does not enable controlled state switching between the green and blue regions. Therefore, the external input is instead injected as a parallel current source $I_{in}(t)$ into the first RTD, leading to the following system of differential equations for the driven case.
\begin{flalign}
    \dot{V}_1 &= \mu^{-1} (I - f(V_1) + I_{in}(t)) \\
    \dot{V}_2 &= (\mu \kappa)^{-1} (I - f(V_2)) \\
    \dot{I} &= \mu (V_0 - V_1 - V_2 - RI)
\end{flalign}
In \panelref{fig:Panel5}{c}, the upper graph shows the pulsed input time series $I_{in}(t)$ that is injected into the first RTD to control the system's state. The voltage responses are shown in the middle and lower graphs. The simulations are initialized in the origin $(V_1, V_2, I)=(0,0,0)$. The time series illustrate that, starting in the lower-right fixed point, the pulses induce transitions to the center and upper-left fixed points in both directions. Immediately before leaving or approaching a fixed point, the system exhibits short transient oscillations. If the system is in either the upper or lower fixed point and additional positive or negative pulses are applied, it exhibits a short excitation but relaxes back to the same stable fixed point. If no further pulses are applied, the system remains in its current stable state, demonstrating memory behavior. \\
The shown behaviour highlights the potential of the series-coupled RTD system as a multistate memory element for neuromorphic hardware. RTDs are commonly studied as artificial neurons exploiting excitability and oscillatory dynamics \cite{OWE26}. For single RTDs, it was shown in \cite{DON24c} that they can exhibit a spiking flip-flop memory consisting of one stable and one oscillatory state. The present analysis extends this concept to a series-coupled RTD system, demonstrating memory based on switching between multiple coexisting stable states.

\section{CONCLUSION}\label{sec:conclusion}

This work investigated series-coupled RTDs in an electrical circuit, resulting in an extended model based on the single-RTD system. Numerical simulations and path continuation methods were employed to analyze the resulting system of differential equations. While the coupled system retains several characteristics of a single RTD, it exhibits new phenomena, including multistability, additional bifurcations, and antisymmetric solutions. Distinct operating regimes in parameter space were identified, and regions suitable for possible neuromorphic operation were determined.

As a consequence of the observed multistability, including both bistable and tristable regimes depending on the bias voltage, the coupled RTD system can operate as a memory element. In particular, external current pulses can induce transitions between coexisting stable states. At $V_0=\qty{6}{V}$, switching between three temporally stable states was demonstrated, thereby enabling tristable memory operation.

By exploiting the separation of time scales between the fast voltage dynamics and the slow current dynamics, it was shown that the equilibrium solutions are restricted to projections of a critical manifold determined by the RTD input–output characteristics. Furthermore, additional limit-cycle branches were found to emerge as the bias voltage $V_0$ is varied. Two-parameter continuation revealed regions of both oscillatory dynamics and stable fixed-point behavior. For the case of two series-coupled RTDs, symmetric and antisymmetric configurations were compared, demonstrating that the antisymmetric configuration supports a richer variety of oscillatory regimes.

The analysis was subsequently generalized to a system of $N$ coupled RTDs revealing that an N-RTD system exhibits $S_N$ symmetry.
All results were obtained for a specific RTD characterized by the parameter set used throughout this work. Although the locations of the operating regimes may shift quantitatively for different RTD designs, the qualitative bifurcation structure is expected to remain unchanged.

\section*{Author contributions}
J.W. did the simulations, analysis of the data, prepared the figures and the manuscript. K.L. and J.J. supervised the work. All authors discussed and wrote the manuscript.
\section*{Acknowledgements}\label{sec:acknowledgements}
This work is supported by the EU Pathfinder Open project 'SpikePro' (Grant ID 101129904).

\bibliography{bibliography}

\newpage
\appendix
\onecolumngrid
\appendix
\renewcommand{\thefigure}{A.\arabic{figure}}
\renewcommand{\theHfigure}{A.\arabic{figure}}
\setcounter{figure}{0}
\setcounter{figure}{0}
\newcommand{\apanelref}[2]{%
  Fig.~\hyperref[#1]{\ref*{#1}(#2)}%
}
\section{Single RTD dynamics}\label{sec:singleRTDdyn}

\begin{figure}[ht]
    \centering
    \includegraphics[width=\textwidth]{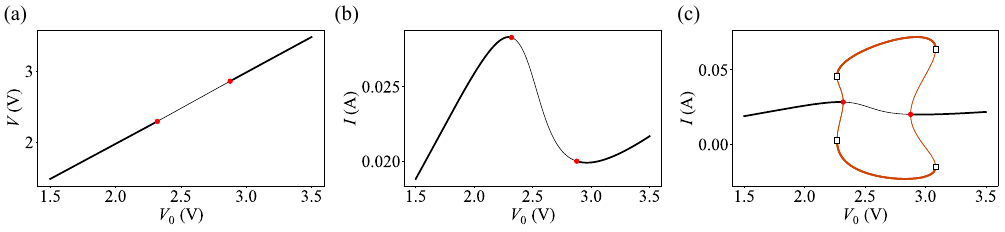}
    \caption{Bifurcation diagrams for the single RTD with $R=\qty{1}{\ohm}$ and $\mu = \qty{0.05}{\per\ohm}$. Thin(thick) lines correspond to unstable(stable) equilibria and unstable(stable) limit cycle oscillations. Red points indicate Andronov-Hopf bifurcations that occur on the fixed-point branches for voltage (a) and (b) current. (c) Same as (b) with limit cycle solutions in red. White squares indicate limit cycle folds.}
    \label{fig:Panel_singleRTD_bifurcations}
\end{figure}

The dynamics of single RTDs were already investigated in detail by Ortega-Piwonka et al.~in \cite{ORT21,ORT21a}. For the single RTD, the dimensionless system of differential equations read
\begin{gather}
    \mu \dot{V} = I-f(V) \\
    \mu^{-1}\dot{I} = V_0 - V - RI
\end{gather}
with the fixed-point equations
\begin{gather}
    f(V) - I = 0\\ 
    V + RI - V_0 = 0
\end{gather}
which can be summarized by the following nonlinear equation    $ V_0 = V + Rf(V)$.
In this case, the critical manifold is simply $\mathcal{C}_0 = \{(V, I): I=f(V)\}$ and the equilibrium branch (black) in \panelref{fig:Panel_singleRTD_bifurcations}{b} shows also merely a projection of the critical manifold. Additional branches do not appear due to the reduced number of equations and the missing symmetry as could be seen in the series-coupled model. As can be seen in Fig.~\ref{fig:Panel_singleRTD_bifurcations}, the equilibrium branch exhibits two Andronov-Hopf bifurcations, in which the real parts of the two eigenvalues cross the axis and the fixed point becomes an unstable node from the earlier stable node. In \apanelref{fig:Panel_singleRTD_bifurcations}{c}, the continuation of the limit cycle from the Andronov-Hopf bifurcations on is shown as orange line while white squares indicate limit-cycle folds.

\begin{figure}[b]
    \centering
    \includegraphics[width=\linewidth]{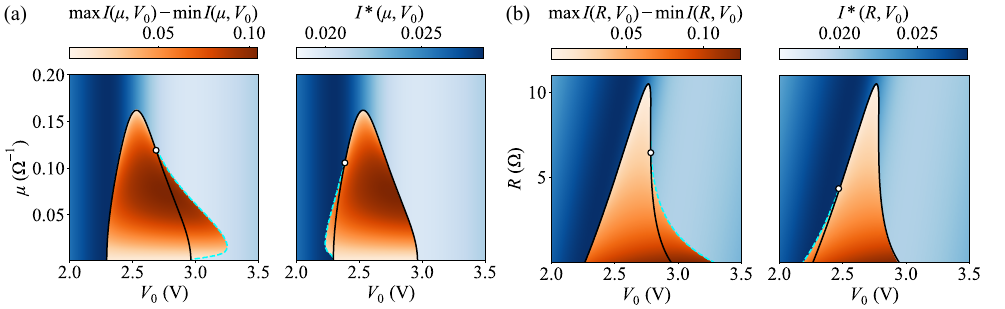}
    \caption{2D bifurcation diagrams for a single RTD in (a) in $\mu$-$V_0$ and in (b) in $R$-$V_0$ parameter space for upward (left) and downward (right) scan. Fixed point current values are indicated in blue and oscillations amplitudes in orange. Black curves represent continued Andronov-Hopf bifurcations and dashed cyan curves continued limit cycle folds. Degenerate Andronov-Hopf points are indicated by white dots and mark the emergence of the limit-cycle folds.}
    \label{fig:Panel_singleRTD_osc}
\end{figure}

The oscillating dynamics, which are born at the Andronov-Hopf bifurcations can be seen in Fig.~\ref{fig:Panel_singleRTD_osc} (orange regions encoding the peak-to-peak amplitude of limit cycle oscillations). At the positions where the Hopf bifurcation (Solid black line) is sub-critical, the system is multistable and the oscillations disappear at the limit-cycle fold (dashed turquoise line) if scanned from the oscillation region outwards. Blue regions indicate the current value observed at the fixed point solution. In \apanelref{fig:Panel_singleRTD_osc}{a} these regions in $\mu$-$V_0$ parameter space can be seen for $R=\qty{1}{\ohm}$ in a $V_0$ forward (left) and backward scan (right). This behavior can easily be explained by considering \apanelref{fig:Panel_singleRTD_bifurcations}{c}. In the forward scan, the system starts on the equilibrium branch and evolves until the branch becomes unstable in the Andronov-Hopf bifurcation. Then the system jumps on the stable limit-cycle branch and evolves until it becomes unstable in the right limit-cycle fold and jumps right back on the equilibrium branch. In the backward scan, the bordering side of the Andronov-Hopf bifurcation and the limit-cycle fold switches because the system approaches the continued limit-cycle branches from the right side in $V_0$.
Similar behavior can be seen for the $R$-$V_0$ scan, despite the oscillating region exhibits a slightly different shape.

\begin{figure}[ht]
    \centering
    \includegraphics[width=0.5\linewidth]{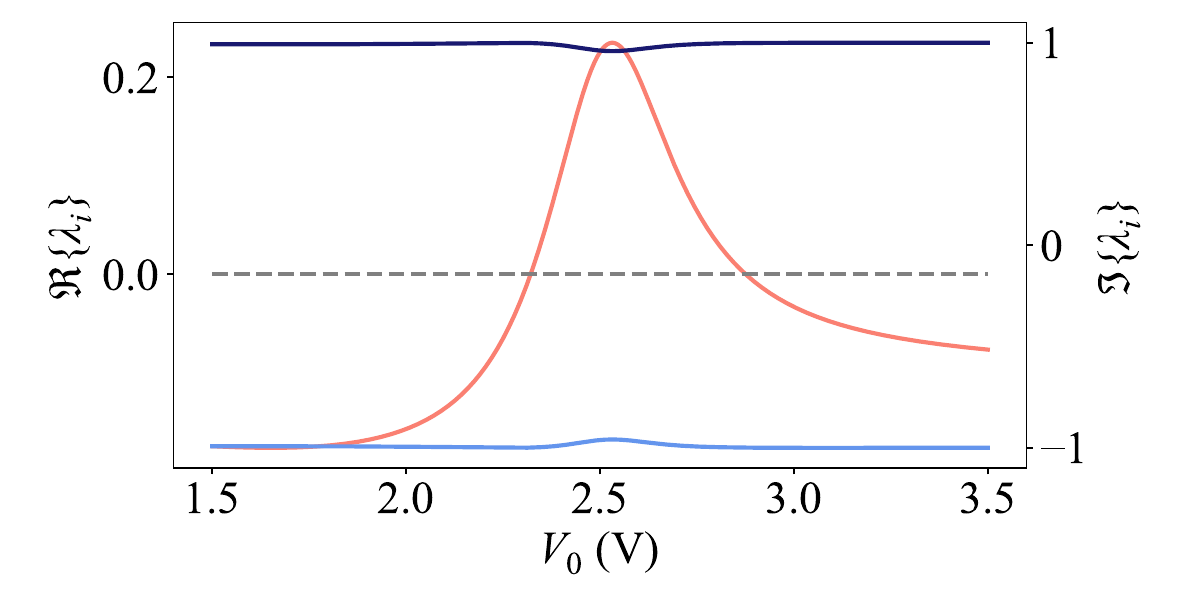}
    \caption{Eigenvalues of the single RTD for $R=\qty{1}{\ohm}$ and $\mu=\qty{0.05}{\per\ohm}$. The red line indicates the real parts of the eigenvalues of $\uu{\bm{J}}$ and the blues lines the imaginary parts. The dashed gray line indicated the zero line of the real part to visualize the zero crossings.}
    \label{fig:Panel_singleRTD_evalues}
\end{figure}

The Jacobian of the system reads
\begin{gather}
    \uu{\bm{J}} = \left(\begin{array}{cc}
        -\mu^{-1}f^{(1)}(V) & \mu^{-1} \\
        -\mu & -\mu R
    \end{array}\right)
\end{gather}
and the eigenvalues of the single RTD are given by
\begin{gather}
    \lambda_{\pm} = -\frac{1}{2}\left[\frac{f^{(1)}(V)}{\mu} + \mu R\right] \pm \frac{1}{2}\sqrt{\left[\frac{f^{(1)}(V)}{\mu} - \mu R\right]^2 - 4}
\end{gather}
based on the characteristic polynomial of $\uu{\bm{J}}$. As shown in Fig.~\ref{fig:Panel_singleRTD_evalues}, the real parts of the eigenvalues are degenerate and the imaginary parts also nearly constant. The zero-crossing of the double eigenvalues indicates the Andronov-Hopf bifurcation in Fig.~\ref{fig:Panel_singleRTD_bifurcations}.

\section{Eigenvalues of series-coupled RTDs}\label{sec:evalues}

\begin{figure}[ht]
    \centering
    \includegraphics[width=\textwidth]{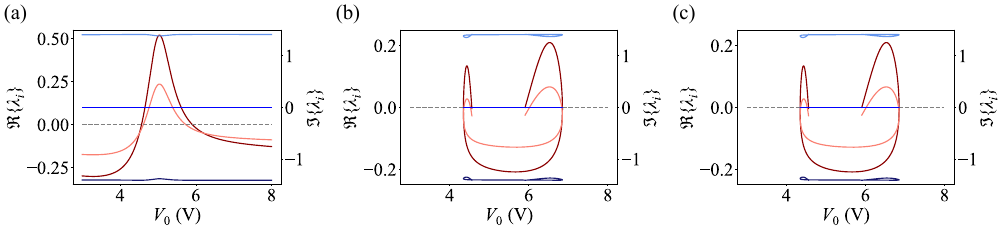}
    \caption{Eigenvalues of (a) main branch, (b) lower side branch and (c) upper side branch. Real parts are indicated in red and imaginary parts in blue. The light red curve represents two degenerate eigenvalues.}
    \label{fig:Panel_evalues}
\end{figure}

The following description refers to the bifurcation diagram shown in \panelref{fig:Panel3}{a} in the main text. The corresponding eigenvalues of the symmetric branch can be seen in \apanelref{fig:Panel_evalues}{a} and of the lower antisymmetric branch in \apanelref{fig:Panel_evalues}{b} and of the upper antisymmetric branch in \apanelref{fig:Panel_evalues}{c}. Zero crossings of the dark red curve for the symmetric branch correspond to the pitchfork bifurcation and of the light red curve to Andronov-Hopf bifurcations. Similar characteristics can be seen for the upper and lower antisymmetric branch where single zero crossing correspond to saddle-node bifurcations and double zero crossing again to Andronov-Hopf bifurcations. Furthermore, it is notable that the imaginary parts show a mirror symmetry with respect to the $\Im\{\lambda_i\} = 0$ line, which supports the interpretation of the Andronov-Hopf bifurcations.

\section{Bifurcations for limit-cycle overlap}\label{sec:app_trans}
\begin{figure}[ht]
    \centering
    \includegraphics[width=\textwidth]{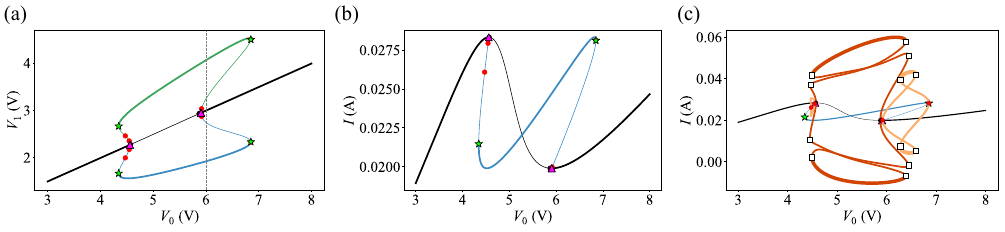}
    \caption{Bifurcation diagrams for series-coupled RTDs with $R=\qty{1}{\ohm}$, $\mu = 0.02\ \unit{\ohm}^{-1}$ and $\kappa=1.1$.}
    \label{fig:Panel_bifurcations}
\end{figure}

As it was already pointed out in Sec.~\ref{sec:dynamics}, especially with respect to \panelref{fig:Panel4}{a}, the system switches from the symmetric limit cycle to the antisymmetric limit cycle for low enough $\mu$. This means, that the continued limit cycles should overlap for small enough $\mu$, which can be seen in the bifurcation diagrams shown in Fig.~\ref{fig:Panel_bifurcations}. These show the symmetry-broken state $\kappa = 1.1$ for $R=\qty{1}{\ohm}$ and $\mu=\qty{0.02}{\per\ohm}$. The Andronov-Hopf and pitchfork bifurcation are shifted along the equilibrium branch while the branches itself are invariant under changes of $\mu$. The stable ranges of the limit-cycle branches at approximately $V_0=\qty{6}{\volt}$ overlap each other, resulting in the two overlapping oscillating region in \panelref{fig:Panel4}{a} that share a border.

\section{Transversality condition for pitchfork bifurcations on the symmetric branch}
The normal form of the symmetry-breaking pitchfork bifurcation is
\begin{gather}
\dot{a} = ra + \beta a^3 + \gamma a^5.
\end{gather}
A necessary condition for the occurrence of the bifurcation is that the coefficient of the linear term vanishes at the critical parameter value $V_0=V_{0,p}$,
\begin{gather}
    r(V_{0,p})=0,
\end{gather}
which is equivalent to $f^{(1)}(s)|_{s=s_p}=0$ with $s_p = s(V_{0,p})$. However, this condition alone does not ensure that the eigenvalue associated with the antisymmetric mode changes sign as $V_0$ passes through $V_{0,p}$. For this purpose, the transversality condition
\begin{gather}
    \left.\frac{dr}{dV_0}\right|_{V_0=V_{0,p}} \neq 0
\end{gather}
must also be satisfied. Using $r(V_0)=-\mu^{-1}f^{(1)}(s(V_0))$, the chain rule yields
\begin{gather}
    \frac{dr}{dV_0} = -\mu^{-1}\frac{d^2f}{ds^2}\frac{ds}{dV_0}.
\end{gather}
Furthermore, differentiating the equilibrium relation $V_0=2s+Rf(s)$ gives
\begin{gather}
    \frac{ds}{dV_0} = \frac{1}{2+R\frac{df}{ds}}.
\end{gather}
Evaluating this expression at the bifurcation point, where $f^{(1)}(s)|_{s=s_p} = 0$, leads to
\begin{gather}
    \left.\frac{dr}{dV_0}\right|_{V_0=V_{0,p}} = -\frac{1}{2\mu}\frac{d^2f}{ds^2}\Bigg|_{s=s_p} \neq 0.
\end{gather}
Hence, the transversality condition is equivalent to requiring that the second derivative of the nonlinear characteristic does not vanish at the bifurcation point.

\section{Transversality condition for Andronov-Hopf bifurcations on the symmetric branch}\label{sec:transversality_condition}
At an Andronov–Hopf bifurcation, the real part of the critical eigenvalue pair must cross the imaginary axis rather than merely touch it \cite{STR15,KUZ23}. Therefore the transversality condition
\begin{gather}
    \frac{d}{dV_0} \Re\{\lambda(V_0)\}\Bigg|_{V_0=V_{0,\text{AH}}} \neq 0
\end{gather}
is used, which reduces for two dimensional systems to
\begin{gather}
    \frac{d\text{tr}(\uu{\bm{J}})}{dV_0}\Bigl|_{V_0=V_{0,\text{AH}}}  \neq 0
\end{gather}
because $\Re\{\lambda\} = \text{tr}(\uu{\bm{J}})/2$ \cite{STR15,KUZ23}. Since the symmetric subsystem $(s,I)$ of $(s,a,I)$ is used for the treatment of the symmetric equilibrium branch, the trace of the Jacobian reads
\begin{gather}
    \text{tr}(\uu{\bm{J}}) = -\frac{f^{(1)}(s)}{\mu} - \mu R
\end{gather}
and the derivative with respect to $V_0$ becomes
\begin{gather}
    \frac{d\text{tr}(\uu{\bm{J}})}{dV_0}\Bigl|_{V_0 = V_{0,\text{AH}}} = -\frac{f^{(2)}(s)|_{s=s_{\text{AH}}}}{\mu}\frac{ds}{dV_0}\Bigl|_{V_0=V_{0,\text{AH}}}.
\end{gather}
In the following, we use the fixed-point equations of the symmetric subsystem $I=f(V)$ and $V_0 = 2s + Rf(s)$ and get
\begin{gather}
    \frac{dV_0}{ds} = 2 + Rf^{(1)}(s) \Leftrightarrow \frac{ds}{dV_0} = \frac{1}{2 + Rf^{(1)}(s)}.
\end{gather}
Using the Hopf condition $f^{(1)}(s)|_{s=s_{\text{AH}}}=-\mu^2 R$ yields
\begin{gather}
    \frac{ds}{dV_0}\Bigl|_{V_0 = V_{0,\text{AH}}} = \frac{1}{2-\mu^2 R^2}.
\end{gather}
Since $\text{det}(\uu{\bm{J}})>0$ ensures that $2-\mu^2R^2$ in the denominator, the transversality conditions reduces to $f^{(2)}(s)|_{s=s_{\text{AH}}} \neq 0$.

\end{document}